\documentclass[12 pt,twoside,english]{article}

\usepackage{amssymb}
\usepackage{amsmath}
\usepackage[T1]{fontenc}
\usepackage[latin10]{inputenc}
\usepackage[a4paper]{geometry}
\usepackage{graphicx}
\usepackage{setspace}
\usepackage{esint}
\makeatletter
\newcommand{\lyxaddress}[1]{
	\par {\raggedright #1
	\vspace{1.4em}
	\noindent\par}
}

\makeatother

\usepackage{babel}
\begin{document}
\title{Spatial autocorrelation and the dynamics of the mean center of COVID-19 infections in Lebanon}
\author{Omar El Deeb $^{ a,b}$}
\date{October 19, 2020}
\maketitle

\lyxaddress{\begin{center}
\emph{$^{a}$Lebanese University, Faculty of Technology - Branch II, Lebanon}\\
\emph{$^{b}$Lebanese International University, Mathematics and Physics Department, Lebanon}\\

\par\end{center}}

\begin{abstract}
In this paper we study the spatial spread of the COVID-19 infection in Lebanon. We inspect the spreading of the daily new infections across the 26 administrative districts of the country, and implement Moran's $I$ statistics in order to analyze the tempo-spatial clustering of the infection in relation to various variables parameterized by adjacency, proximity, population, population density, poverty rate and poverty density, and we find out that except for the poverty rate, the spread of the infection is clustered and associated to those parameters with varying magnitude for the time span between July (geographic adjacency and proximity) or August (population, population density and poverty density) through October. We also determine the temporal dynamics of geographic location of the mean center of new and cumulative infections since late March. The results obtained allow for regionally and locally adjusted health policies and measures that would provide higher levels of public health safety in the country.
\end{abstract}

Keywords: COVID-19; Spatial autocorrelation; Mean center of infection; Lebanon; Health safety.

\maketitle

\section{Introduction}

The spread of COVID-19 pandemic has practically affected the entire planet, and created enormous challenges on every aspect of human life and organization, starting with the health sector and with far reaching consequences on the economy, education, sports, transportation and politics. Since the first cases were registered in Wuhan, China in December 2019 \cite{WHO}, the global spatial dynamics of the infection have been changing as the disease swiftly moved towards the West \cite{Mathys1} into Europe then into the United States, South America, and eventually to the whole world, with nearly 38.1 million cases and 1.1 million deaths registered until October 12, 2020 \cite{Worldometer}.

Given the global geographic spread of the virus and the local wide spread in many countries, and the nature of the transmission of the virus, it is important to understand the spatial mechanisms of this spread and its dependence on proximity, demographics and social characteristics of infected areas. Spatial analysis provides a better understanding of the routes of transmission of infections \cite{Meng1}, consequently, it allows the decision-makers to draft and implement effective health and mitigation measures to reduce risks associated with the pandemic.

In Lebanon, the first case was registered on February 21, 2020 \cite{MOPH} and by October 12, $54624$ cases and $466$ deaths were registered \cite{DRM}. The first few weeks witnessed a relatively rapid increase but it sharply declined as a result of the strong mitigation measures enforced by the beginning of March. The lift of the international travel ban and the partial easing of measures led to the revival of higher spread rates since July. Only $1788$ cases were registered by July 1, 2020 before a sharp rise from July through October. The cases were mainly concentrated in Beirut, its suburbs and its neighboring areas in Mount Lebanon. On August 4, a huge explosion rattled the port of Beirut and destroyed thousands of houses and buildings in the surrounding areas. People were rushed into hospitals, with thousands of injuries recorded on that day \cite{BBC}. On such a horrible incident, hundreds of volunteers and civil defense teams were involved in rescue work for several days. The social distancing measures were largely neglected in such an emergency situation. The spread accelerated in the upcoming weeks, with sharp rise in Beirut and its surroundings and with a national widespread reaching all regions and major towns and cities \cite{Science}.

\begin{figure}[t]
\hspace*{-2cm}
\begin{center}
\includegraphics[scale=0.8]{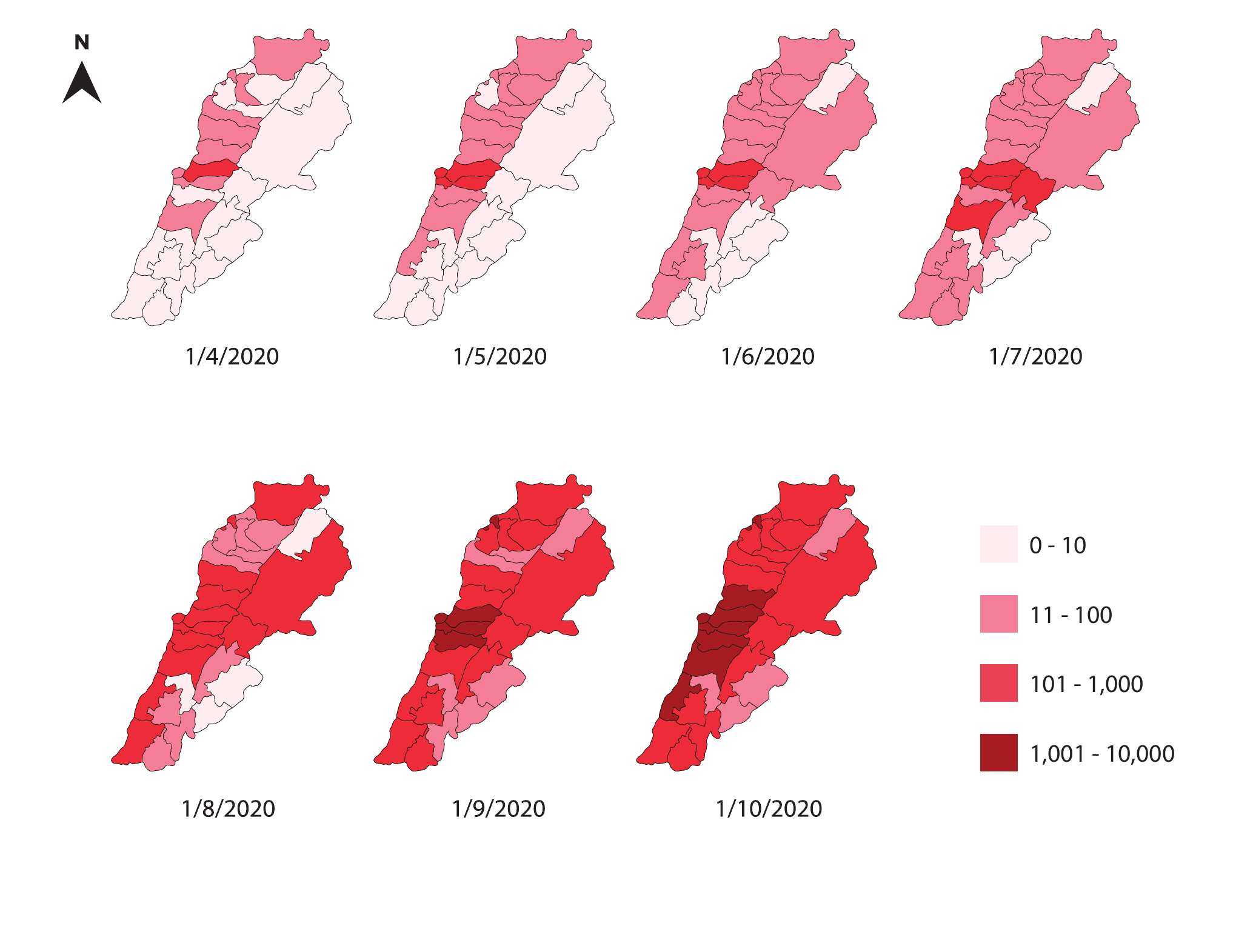}
\caption{A monthly map of the regional cumulative number of infections in Lebanese districts between April and October 2020.}\label{fig:1}
\end{center}
\end{figure}

\emph{Related Literature}: Spatial autocorrelation is the statistical analysis of data studied in space or in space-time aiming for the identification and estimation of spatial processes \cite{CliffOrd1, Anselin}. It has been implemented to study and analyze the spread of various diseases and infections including cancer, diabetes, SARS, influenza virus, COVID-19, etc... \cite{Glick, Hipp, SARS, PLOS, Kuwait}. Recent studies also inspected the effect of city size, population, transportation systems and demographics on the disease spread and its mortality rate \cite{PLOS1, Urban, Xie, Safety1, Safety2}.   The determination of the mean center of a population (centroid) was discussed in \cite{Aboufadel, Bachi, Barmore} and extending the concept to the determination of the mean center of wealth and infections allowed for a spatial analysis of the temporal dynamics of wealth distribution, economic growth and infectious diseases \cite{Mathys2}. The dynamics of the outbreak of COVID-19 in Lebanon and its reproduction number dynamics were studied in \cite{DeebJalloul1, DeebJalloul2, Mourad}.

In this paper, we study the clustering and spatial progression of new infections in Lebanon by applying the methods of spatial autocorrelation with different model parameterizations of geographic, demographic and social variables including  adjacency, proximity, population, population density, poverty rate and poverty density. Locating the mean center of the epidemic spread as a function of time is used to analyze the temporal geographic development of the spread. The obtained results provide a solid basis for the concerned policy makers to draw well-grounded and scientifically based local and regional measures that would contribute to controlling the infection spread.

 The paper is organized as follows: in section 2 we introduce the implemented analytic mathematical and statistical methods and tools. Results are presented and discussed in section 3, and section 4 concludes the paper.

\section{Analytic methods and tools}

\subsection{Moran's I index}

Moran's $I$ index is an inferential statistic used to measure the spatial autocorrelation based both on  locations and feature values simultaneously. It is defined as \cite{CliffOrd1}:
\begin{equation}
 I=\frac{N\Sigma_{ij}W_{ij}(X_{i}-\bar{X})(X_{j}-\bar{X})}{\Sigma_{ij}W_{ij}\Sigma_{i}(X_{i}-\bar{X})^{2}} 
\end{equation}

where $W_{ij}$ represents different types of adjacency between
region $i$ and region $j$, corresponding to different models of
infectious spread. $N$ is the number of regions under consideration
and $X_{i}$ represents the number of new daily infections
in district $i$. $\bar{X}$ is the average number
of new daily infections per region, and it is given by $\bar{X}=\frac{\Sigma_{i}X_{i}}{N}$. 
The numerical outcome of $I$ falls between $-1$ and $1$ and it indicates whether a distribution is dispersed, random or clustered. A value of $I$ close to $0$ indicates a random distribution, while positive values  indicate clustered spatial distribution and negative values indicate dispersion. Larger values of $|I|$ nearer to $1$ mean stronger clustering (positive $I$) or stronger dispersion (negative $I$).

The $z_{I}$-score associated to this statistic is defined by:

\begin{equation}
z_{I}=\frac{I-E[I]}{\sqrt{V[I]}}
\end{equation}

where the expected value $E[I]$ and the variance $V[I]$ are defined in the Appendix. The $z$-score or the corresponding $p$-value of the statistic are used to reject the null hypothesis and eliminate the possibility of a random pattern leading to the obtained value of the Moran $I$ statistic.

In this paper, we take a $95\%$ confidence level corresponding to $|z_{I}|>1.96$ or equivalently to $p<0.05$ in order to confirm the outcome of clustering or dispersion of our spatial data indicated by $I$. In this case we say that the $p$-value is statistically significant, and based on the value of $I$ we can determine the pattern of the distribution. 

We consider a model with six different cases of parameterization of the adjacency matrix $W_{ij}$ corresponding to geographic adjacency (case I), proximity (case II), population (case III), population density (case IV), poverty rate (case V) and poverty density (case VI). Table \ref{table:1} summarizes relevant data from the Lebanese districts.

In  casel I, we take $W_{ij}=1$ for districts sharing common borders and $W_{ij}=0$ otherwise, while in case II we determine $W_{ij}=\frac{1}{d_{ij}}$ where $d_{ij}$ is the driving distance between the administrative centers of regions $i$ and $j$. Those two cases study the effect of administrative adjacency and the distance proximity of different districts on the geographic clustering of new infections in Lebanon. 

In case III and case IV, we apply the methods used in (\cite{Meng1, Kang}) to analyze the effects of population and population density on the spread of the disease since the virus is carried by people and its spread is supposed to be related to their population and interaction. We sort the districts by the number of their residents (obtained from \cite{CAS})  and then by the density of their residents relative to their areas. In these two cases, districts of consecutive populations and population densities  are assigned a factor of $W_{ij}=1$, and $W_{ij}=0$ otherwise. This provides a statistic about the clustering of infections according to population and population density respectively.

Lastly, in cases V and VI, we introduce  new parameters, namely the poverty rate and the poverty density in different districts and we analyze their effect on infection clustering. We sort the distritcs by their rates of poverty  and poverty density  \cite{CAS} and assign  $W_{ij}=1$ for regions of consecutive order of poverty rate or poverty density, and $W_{ij}=0$ otherwise, in a similar methodology to cases III and IV in order to infer the effect of poverty rate and density on the geographical patterns of infection spread.

\subsection{Mean center of infection}

The mean center of infection (henceforth MCI) is a geographic location that represents the weighted mean of the positions of infected individuals on the surface of Earth, assumed to be spherical. Assigning the value of Earth's radius to unity, the two spherical coordinates that determine the unique position of a point are its latitude $\lambda_{i}$ and longitude $\phi_{i}$.  The latitude is a measurement of location north or south of the equator while the longitude is a measurement of location east or west of the prime meridian at Greenwich, UK. 

The Cartesian position vector $\overrightarrow{r_{i}}=(x_{i},y_{i},z_{i})$
is related to spherical coordinates with unit radius by the relations \cite{Coordinates}:

\begin{equation}
\begin{cases}
x_{i}=\cos\lambda_{i}\cos\phi_{i}\\
y_{i}=\sin\lambda_{i}\cos\phi_{i}\\
z_{i}=\sin\phi_{i}
\end{cases}
\end{equation}

We denote the district number of infections (new or cumulative) by $X_{i}$ as defined above, and the Cartesian positions of the administrative centers by $(x_{i},y_{i},z_{i})$ . Then, the Cartesian position of the weighted mean of infections
$\overrightarrow{\hat{r_{i}}}$ is given by:

\begin{equation}
\begin{cases}
\hat{x}=\frac{\Sigma_{i}X_{i}x_{i}}{\Sigma_{i}X_{i}}\\
\hat{y}=\frac{\Sigma_{i}X_{i}y_{i}}{\Sigma_{i}X_{i}}\\
\hat{z}=\frac{\Sigma_{i}X_{i}z_{i}}{\Sigma_{i}X_{i}}
\end{cases}
\end{equation}

As suggested by \cite{Aboufadel}, the precise position on the surface of a sphere can
be determined from the normalized position vector defined by $\overrightarrow{\bar{r_{i}}}=(\bar{x},\bar{y},\bar{z})=\frac{\overrightarrow{\hat{r_{i}}}}{\mid\overrightarrow{\hat{r_{i}}}\mid},$leading
to

\begin{equation}
\begin{cases}
\bar{x}=\frac{\hat{x}}{\sqrt{\hat{x}^{2}+\hat{y}^{2}+\hat{z}^{2}}}\\
\bar{y}=\frac{\hat{y}}{\sqrt{\hat{x}^{2}+\hat{y}^{2}+\hat{z}^{2}}}\\
\bar{z}=\frac{\hat{z}}{\sqrt{\hat{x}^{2}+\hat{y}^{2}+\hat{z}^{2}}}
\end{cases}
\end{equation}

\begin{figure}[t]
\hspace*{-2cm}
\begin{center}
\includegraphics[scale=0.55]{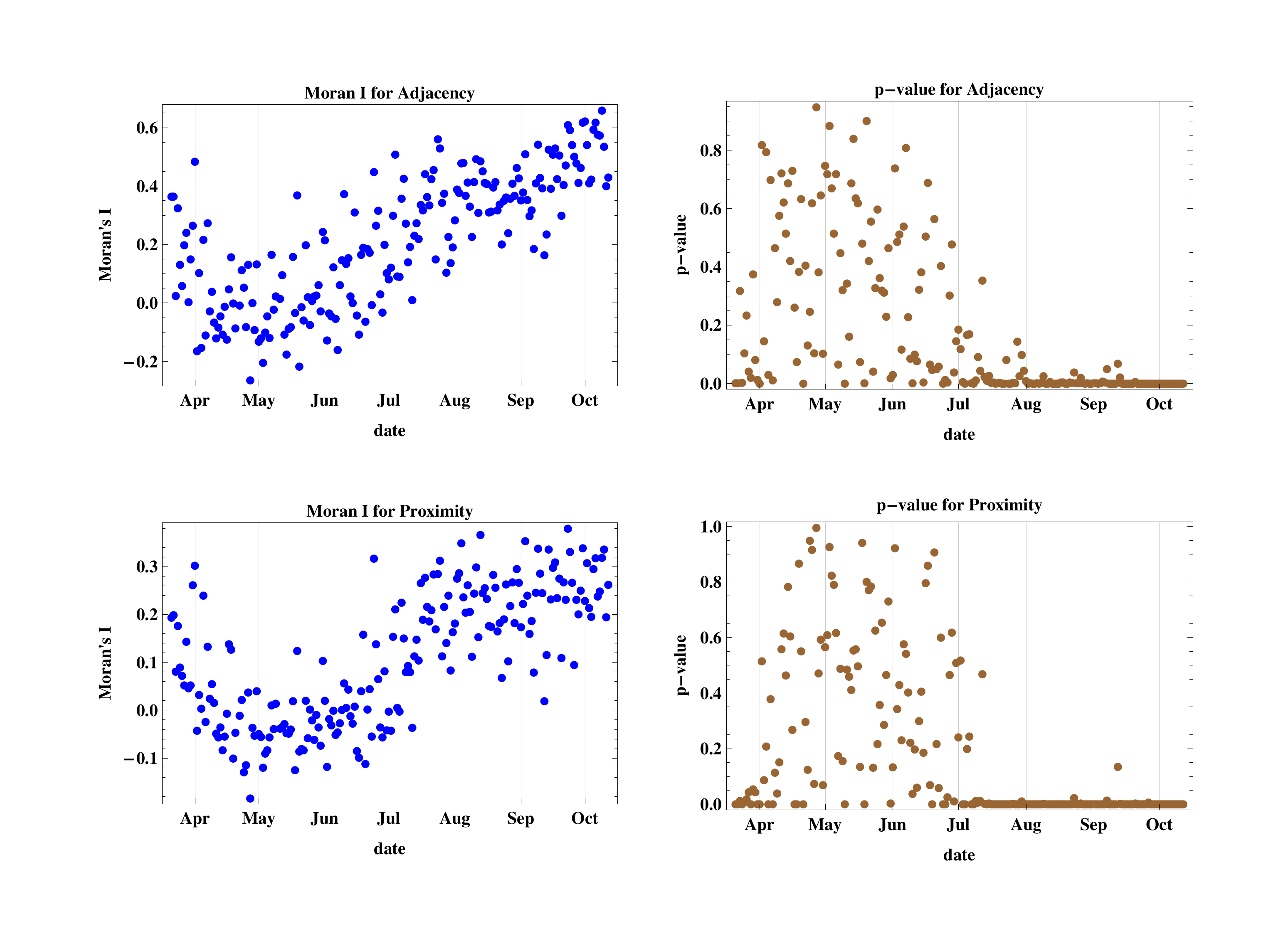}
\caption{The figure shows Moran's I index and its corresponding p-value for cases I and II accounting for adjacency and proximity of new infections registered in Lebanese districts. The test shows that since July 2020, there is clear geographic clustering of infections. }\label{fig:2}
\end{center}
\end{figure}

\begin{figure}[h]
\hspace*{-2cm} 
\begin{center}
\includegraphics[scale=0.55]{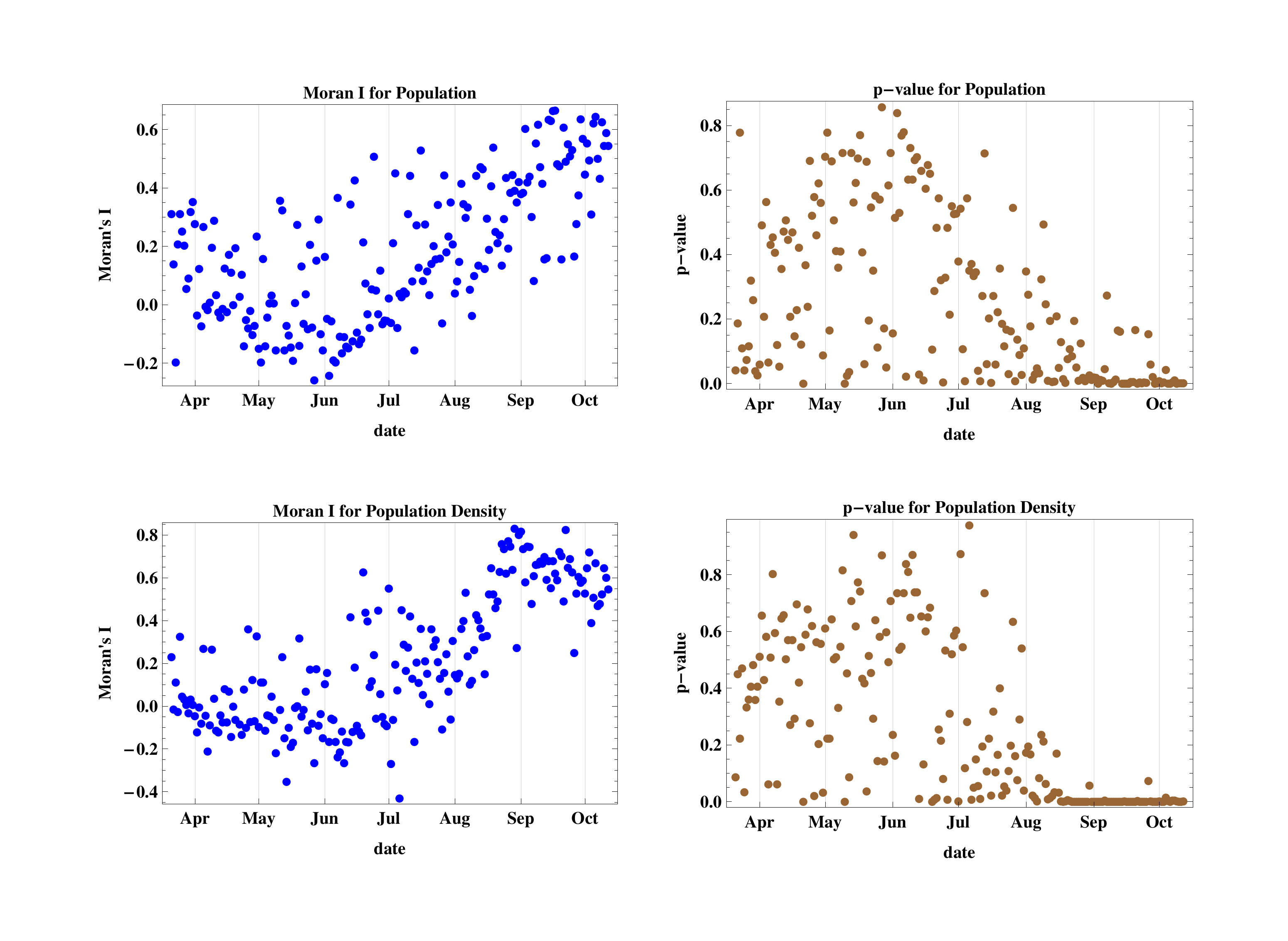}
\caption{Moran's I index and its corresponding p-value for cases III and IV accounting for population and population density of different districts. Population density contributes for strong clustering since August 2020, while the districts' population effect is weaker. } \label{fig:3}
\end{center}
\end{figure}

Consequently, we can recover the spherical position of the mean center of infections by calculating the mean latitude and longitude as:

\begin{equation}
\begin{cases}
\bar{\phi}=\sin^{-1}\bar{z}\\
\bar{\lambda}=\tan^{-1}\frac{\bar{y}}{\bar{x}}
\end{cases}
\end{equation}

The latitude and the longitude can be located and plotted on maps and geographic information systems. We employ the spherical coordinates of geographic locations of the capitals of the $26$ administrative districts in Lebanon and the number of daily and cumulative infections in each region in order to determine the daily MCI accordingly. This provides a tool to analyze the temporal dynamics of the mean geographic spread of the disease.

\section{Results and discussions}

The determination of the Moran's $I$ index and its corresponding p-value for the effect of adjacency and proximity of cases I and II on the clustering of daily new infections of COVID-19 in Lebanon shown in Figure (\ref{fig:2}), leads to the conclusion that since July 2020, there is strong clustering of infections in regions sharing common borders and among nearer regions. There were only few days when new infections were not clustered in adjacent regions, and only one day where distance was not shown to be a detrimental effect in the spatial spread of new cases. The maximum value of Moran's $I$ reached $0.660$ for case I and $0.380$ for case II indicating a high level of geographic clustering of the disease spread since July. The infections before July had a high $p$-value, indicating a high probability for random geographic spread.

\begin{figure}[t]
\hspace*{-2cm}
\begin{center}
\includegraphics[scale=0.55]{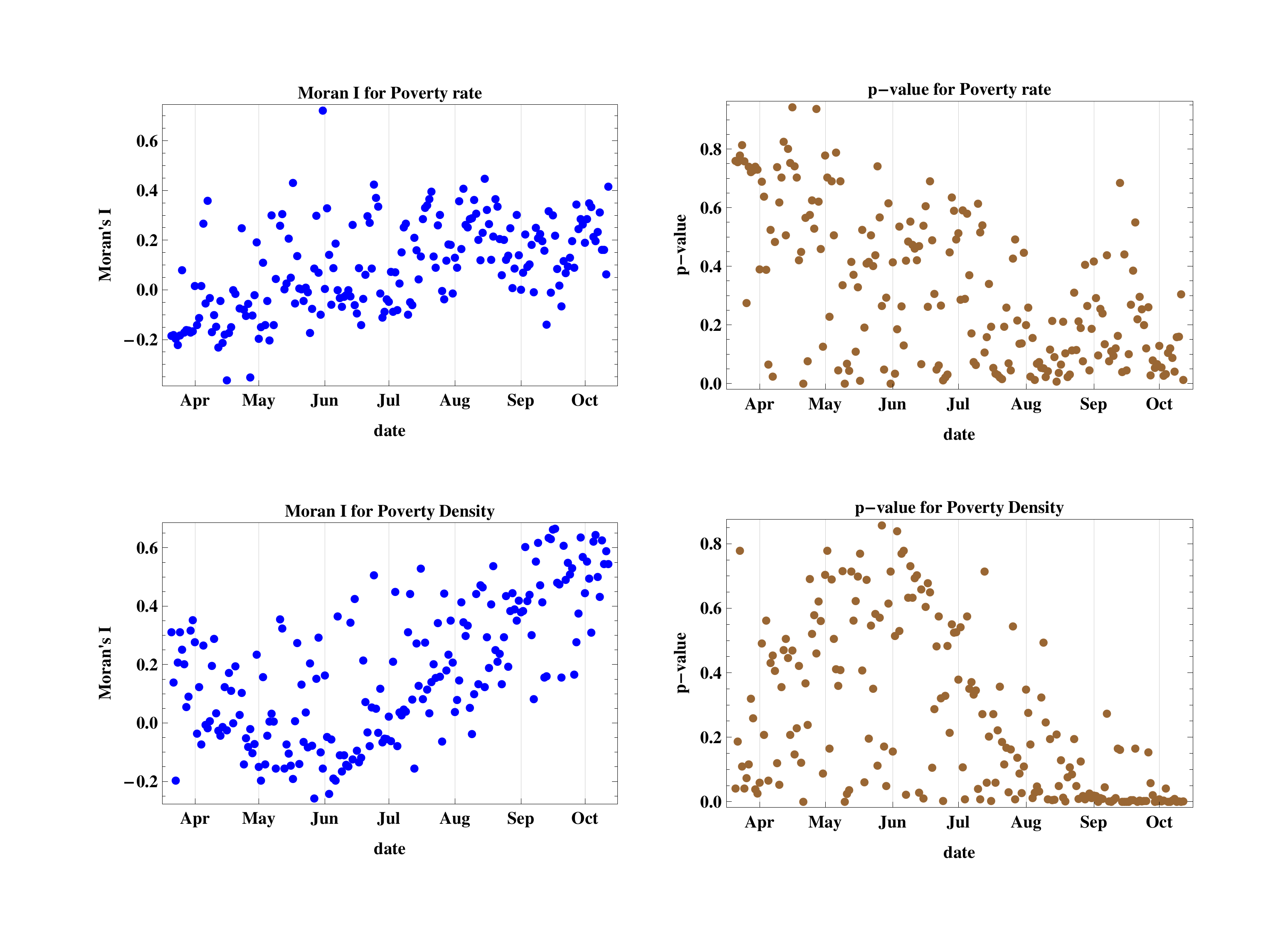}
\caption{Moran's I index and its corresponding p-value for cases V and VI accounting for regional poverty rate and poverty density. The poverty rate is not a decisive factor for spatial spread but poverty density contributes to spatial clustering since the end of August 2020. }\label{fig:4}
\end{center}
\end{figure}

\begin{figure}[t]
\hspace*{-2cm}
\begin{center}
\includegraphics[scale=0.55]{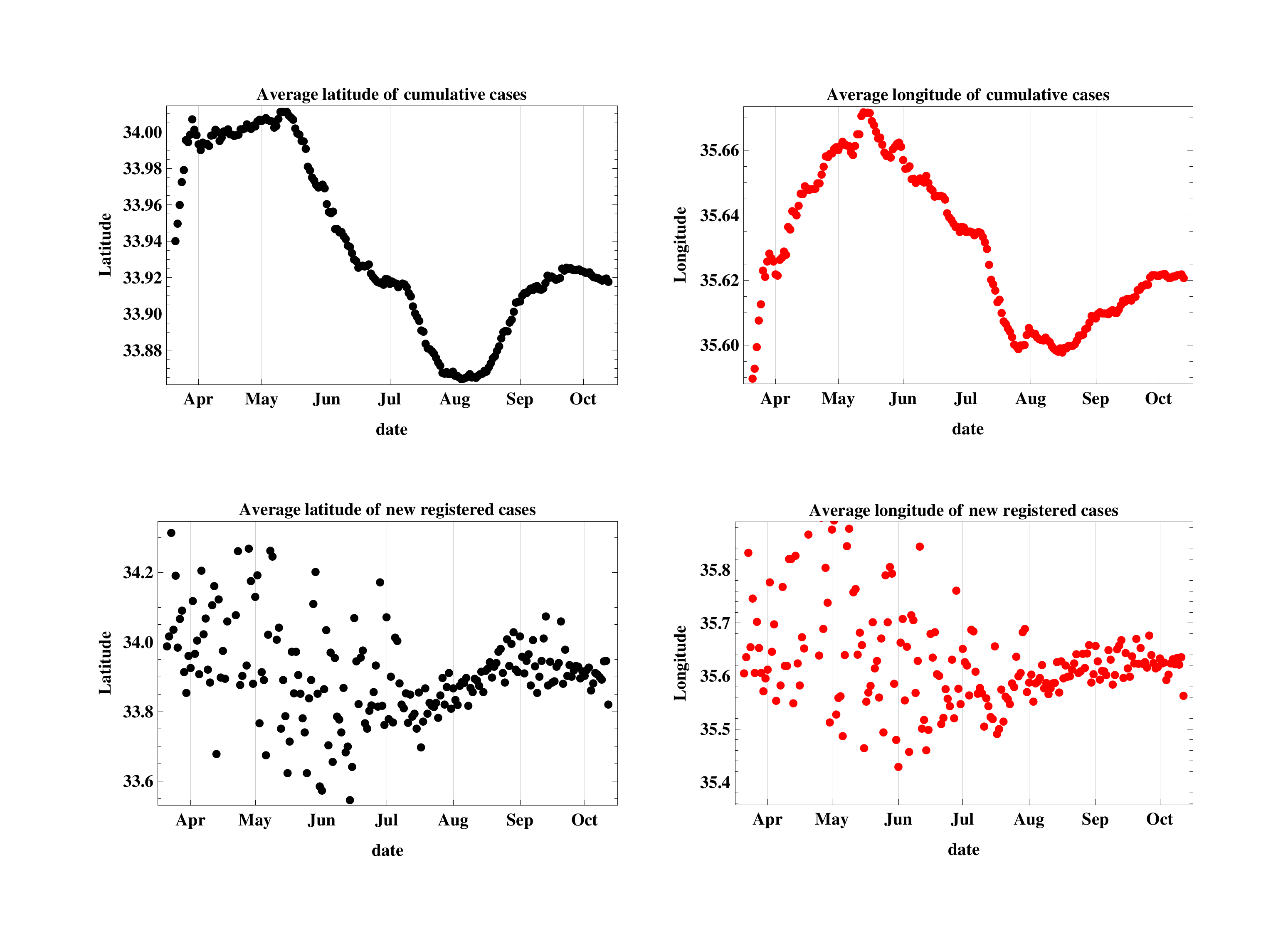}
\caption{This figure is a plot of the latitude and longitude of the weighted geographic center of COVID-19  infections in Lebanon. The upper graphs represent the temporal progression of the cumulative number of infections while the lower graphs represent that of the new daily cases.}\label{fig:5}
\end{center}
\end{figure}

The results of the spatial spread dynamics in relation to population and population density adjacency as shown in Moran's $I$ and $p$-value of cases III and IV depicted in  Figure (\ref{fig:3}) reveal that the spread was not clustered with respect to the regional population until late August 2020, where it started achieving a positive value of $I$ with $p<0.05$ indicating spatial clustering between regions of adjacent population rank, with several days showing a probability of random spread. The maximum attained $I$ was $0.666$. However, the statistics for districts with adjacent rank of population density show very strong spatial clustering since the middle of August with $I$ attaining a maximum value of $0.832$, which is the highest among all six studied cases.

The results of case V  (Figure \ref{fig:4}) show that the spatial spread cannot be attributed to adjacent ranking of poverty rates among the districts since the p-values remain above the $5\%$ level of confidence up until October 2020, hence no spatial clustering occurs. But when we consider the poverty density in case VI, we obtain positive values for Moran's $I$ since the end of August, with $p<0.05$ except for five days. Hence, spatial clustering among regions with adjacent ranking of poverty density occurs. The maximum attained $I$ in this case is $0.666$.

In comparison, we find out that clustering of new infections occurs starting on different dates between July and August for all considered cases except for case V corresponding to district populations. The strongest level of spatial clustering (highest $I$) occurs for model IV of population density after mid-August, while clustering associated to geographic adjacency and proximity (cases I and II) has the longest time span (since early July) and the highest levels of confidence.

\begin{figure}[t]
\hspace*{-2cm}
\begin{center}
\includegraphics[scale=0.65]{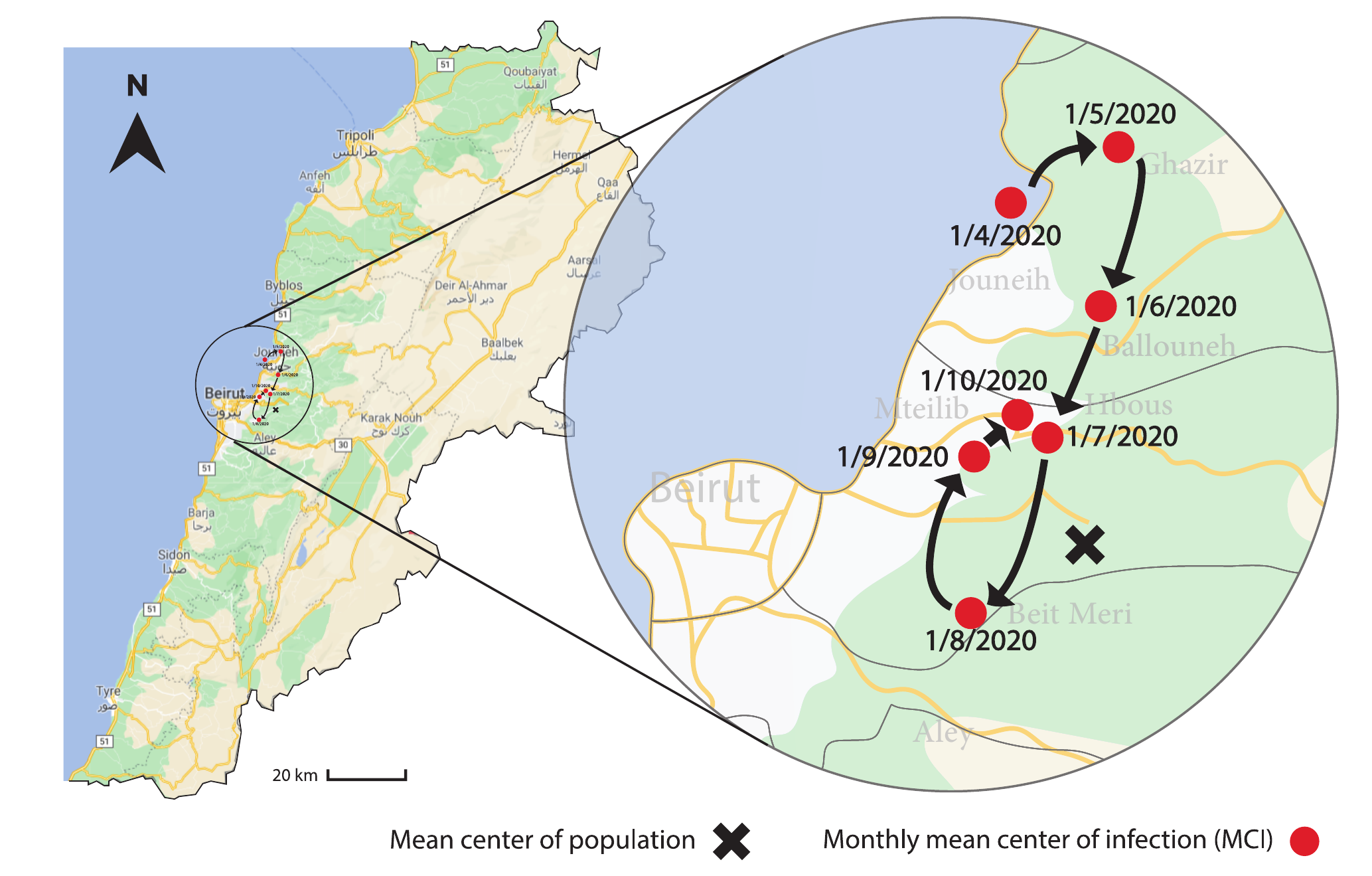}
\caption{The figure shows the variation of the geographic location of the MCI on a monthly basis between April 1 and October 1, 2020, together with the mean center of the population.}\label{fig:6}
\end{center}
\end{figure}

The location of the MCI was determined as a function of time as shown in   Figure (\ref{fig:5}). The mean latitude and longitude of the infection were determined according to the methods described in equation (6). The location of the cumulative MCI is plotted on the geographic map of Lebanon during the same period in  Figure (\ref{fig:6}), together with the mean center of population of the country. It started near the city of Jounieh, north-west of the mean center of population, but it has moved southward since May through August, where it started moving northward again. The location of the MCI of new infections was quite geographically distributed before July as the lower plot of  Figure (\ref{fig:5}) shows, before becoming more homogenous afterwards.

The reproduction number $R$ and the rate of the infection spread correlate with people's mobility \cite{Safety3}. Geographic clustering occurs because people's motion and local travel is higher in their close neighborhoods, especially in a country like Lebanon where with the absence of national public transportation throughout the country \cite{Transports} diminishes nationwide mobility. Higher levels of social interaction among people in dense regions also contribute to the spread of the disease, and this has shown the strongest clustering effect. The study of the spread of the infection allows the relevant authorities to draw appropriate country-specific and regional measures to curb the spread.

\begin{flushleft}

\begin{table}[t]
\begin{raggedright}
{\small{}}%
\caption{The table shows the distribution of the cumulative number of cases among the 26 Lebanese districts on October 12, 2020, with their respective populations, population densities, poverty rates and poverty densities.} \label{table:1}
\smallskip
\begin{tabular}{lc lc lc lc lc lc}
\hline 
{\small{}Region Name} & {\small{}Number of cases} & {\small{}Population } & {\small{}Population denisty} & {\small{}Poverty} & {\small{}Poverty density}\tabularnewline
{\small{}} & {\small{} (Cumulative)} & {\small{} $\times100$} & {\small{}(resident/$km^{2}$)} & {\small{}rate (\%)} & {\small{}(resident/$km^{2}$)}\tabularnewline
\hline 
{\small{}Akkar} & 1171 & 3204 & 418  & 38.4 & 38 \tabularnewline
{\small{}Minieh-Denniyeh} & 723  & 1408  & 389  & 48.6  & 189 \tabularnewline
{\small{}Tripoli} & 4198 & 2438 & 9030 & 31.7 & 2862 \tabularnewline
{\small{}Zgharta} & 854 & 877 & 399  & 25  & 100 \tabularnewline
{\small{}Koura} & 556 & 846 & 489 & 14.3 & 70 \tabularnewline
{\small{}Bcharre} & 115 & 221  & 140  & 13.4  & 19 \tabularnewline
{\small{}Batroun} & 341 & 589 & 212 & 5.5 & 12 \tabularnewline
{\small{}Jbeil} & 954 & 1295 & 301  & 12.7  & 38 \tabularnewline
{\small{}Kesrwan} & 1978 & 2605  & 762 & 18.5 & 141 \tabularnewline
{\small{}Meten} & 6139  & 5110  & 1928  & 20.4  & 393 \tabularnewline
{\small{}Beirut} & 6443  & 3417  & 17258  & 25.5 & 4401 \tabularnewline
{\small{}Baabda} & 7277  & 5538  & 2855 & 26.8  & 765 \tabularnewline
{\small{}Aley} & 3047 & 3008 & 1144  & 29.4  & 336 \tabularnewline
{\small{}Chouf} & 1965  & 2770  & 560  & 24.2  & 135 \tabularnewline
{\small{}Jezzine} & 125 & 321  & 133  & 21.9  & 29 \tabularnewline
{\small{}Saida} & 2472  & 2966  & 1079  & 19.1  & 206 \tabularnewline
{\small{}Nabatieh} & 683  & 1802  & 593  & 28.2  & 167 \tabularnewline
{\small{}Sour} & 1023 &  2557 & 933  & 30.3 & 283 \tabularnewline
{\small{}Bent Jbeil} & 331 & 962  & 364  & 22.9  & 83 \tabularnewline
{\small{}Marjeyoun} & 214 & 740  & 279  & 24.2  & 68 \tabularnewline
{\small{}Hasbaya} & 76 & 287  & 108  & 23.9  & 26 \tabularnewline
{\small{}Rachaya} & 79 & 338  & 62  & 16  & 10 \tabularnewline
{\small{}West Beqaa} & 474  & 864 & 184  & 25.5  & 47 \tabularnewline
{\small{}Zahleh} & 2367 & 1774  & 424  & 37.3  & 158 \tabularnewline
{\small{}Baalback} & 877  & 2146  & 94  & 40.6 & 38 \tabularnewline
{\small{}Hermel} & 86  & 305  & 42  & 47.1  & 20 \tabularnewline
\hline 
\end{tabular} {\small\par}
\par\end{raggedright}

\end{table}
\par\end{flushleft}

\section {Conclusion}

 In this paper we introduced the Moran's $I$ index with its associated $z$-score and $p$-value to study the spatial autocorrelation of registered new infections of COVID-19 in Lebanon. We introduced six different cases of parameterization of the spread related to adjacency, proximity, population, population density, poverty rate and poverty density. We discovered that poverty rate is not statistically relevant to the spatial spread of the disease while geographic bordering, distance between district centers, number and density of residents and poverty density lead to clustering of the disease, with varying strengths and level of confidence since July and August through October. We also introduced methods to determine the geographic coordinates of the mean center of the infection, and determined this center since April 2020, and plotted its variations over time up until October.
The understanding of the spatial, demographic and geographic aspects of the disease spread over time provides an essential basis for the relevant authorities to take more efficient decisions of local and inter and intra-regional measures, thus contributing to increased social and health safety and security in the fight against the pandemic.

\section*{Appendix}

 The expected value of Moran's $I$ statistic is given by: 
\[E[I]=\frac{-1}{N-1}\]

while its variance is defined as: 

 \[V[I]=E[I^{2}]-E[I]^{2}\]
 where

\[E[I^{2}]=\frac{A-B}{\left(N-1\right)\left(N-2\right)\left(N-3\right)\left(\Sigma_{ij}W_{ij}\right)^{2}}\]

and $A$ and $B$ are given by:
\[
A=N\left[2\left(N^{2}-3N+3\right)\Sigma_{ij}W_{ij}^{2}-2N\Sigma_{i}\left(\Sigma_{j}W_{ij}\right)^{2}+3\left(\Sigma_{ij}W_{ij}\right)^{2}\right]
\]

\[
B=\frac{2\Sigma_{i}(X_{i}-\bar{X})^{4}}{\left(\Sigma_{i}(X_{i}-\bar{X})^{2}\right)^{2}}\left[\left(N^{2}-N\right)\Sigma_{ij}W_{ij}^{2}-2N\Sigma_{i}\left(\Sigma_{j}W_{ij}\right)^{2}+3\left(\Sigma_{ij}W_{ij}\right)^{2}\right]
\]
consequently, the $z_{I}$-score is given by $ z_{I}=\frac{I-E[I]}{\sqrt{V[I]}}$.

\section*{Acknowledgements}
The author has no acknowledgements.

\section*{Data Availability}
Raw data was obtained from publically available and cited resources. All used data and codes are available upon request.

\end{document}